# Finding the boson-number distributions in superconductors from photon spectra emitted by supercurrent-carrying rings irradiated with far IR fields

A. I. Agafonov

National Research Center "Kurchatov Institute", Moscow, 123182 Russia

E-mail: aai@isssph.kiae.ru; aiagafonov7@gmail.com

**Abstract.** A theory of the IR-field-induced single-photon generation by the narrow thin-film superconducting rings made of the isotropic *s*-wave pairing type-II superconductors is presented. It is shown that statistical measurements of the energies of photons emitted by the same current-carrying ring prepared initially in the same quantum state, allow to find the number distribution of Cooper pairs in the superconductor.

**Keywords:** superconducting ring, gauge invariance breaking, coherent macroscopic quantum state, photon generation, boson-number distribution

## 1. Introduction

It is generally accepted that the superconducting transition is accompanied by spontaneous breaking of the gauge invariance [1-6]. As a consequence, the superconducting state is characterized by the complex off-diagonal long-range order parameter. Both the amplitude and the phase of the parameter are fixed, but the number of the Cooper pairs is not conserved. The superconducting condensate can be described as the coherent superposition of states $|N>$ with different numbers of the Cooper pairs [2,3,5]:

$$\Psi = \sum_N c_N \Psi_N |N>, \qquad (1)$$

where $c_N^2$ gives the boson-number distribution in the superconductor, and $\Psi_N$ is the many-body wave function which describes all the pairs in the same quantum state with the well-defined phase ($\phi$),

$$\Psi_N \propto \exp(iN\phi). \qquad (2)$$

This spontaneous gauge-symmetry breakage is the basis of the modern theory of superconductivity. Note that the superconducting state in the Bardeen-Cooper-Schrieffer theory can be presented in the form (1)-(2) with the amplitudes $c_N$ peaked up around the average value $\overline{N}$ [7]. The width of this peak $\Delta N$, which is believed to be proportional to $\sqrt{N}$, is the standard deviation of the number of bosons in the condensate or, in other words, the uncertainty in their numbers.

Far as we know, analytical expressions for $c_N$ have still not been theoretically received, and the boson-number distributions in superconductors have not been experimentally studied. Similar to the coherent states in quantum optics [8], it was assumed that the values $c_N^2$ can be presented by the Gaussian distribution [2,3] or the Poissonian statistics [5]. Determination of these $N$ distributions, which can depend on the superconductor size and external parameters such as the temperature and magnetic field, would be a good verification of our knowledge of the superconducting state. Therefore, an experiment for finding $c_N^2$ is seemed to be urgent.

The fact that the superconducting condensate possesses the phase, leads to a number of experimental consequences [7]. One of them is persistent supercurrent that can flow in multiply-connected structures, in particular, in rings in the absence of external magnetic fields. These currents are, in principle, metastable, and can be varied by the quantum jumps, at which the number of the magnetic-flux quanta, trapped in the superconductor, changes by one or several unites (see [9-12] and references therein). Also, the quantized decay of the current states can be due to the inelastic magnetic neutron scattering by superconducting rings [13].

The supercurrent-carrying rings have discrete energy levels. These energies are equal to the ring characteristic energy multiplied by the square of the number of the magnetic-flux quanta trapped into the ring (or the fluxoid number). In turn, the characteristic energy is proportional to the boson number which is characterized by the distribution function in the superconductor.

For the supercurrent state with the given fluxoid number all other states with smaller values of this number are states with lower energies. An external perturbation field is needed to put the superconducting ring into a lower energy state, and the energy difference between the initial and final supercurrent states can be carried away by one or several photons. At the same time, the energy of these photons will depend on the number of bosons. This process, however, is required the simultaneously transition of a large number of the Cooper pairs, the density of which is about of $10^{20}$ cm$^{-3}$. Nevertheless, these external-field-induced quantum jumps of the supercurrent with emission of photons can have a finite probability due to the features of the superconducting condensate state discussed above.

A far-IR radiation can be used as the external perturbation of the condensate in the superconducting rings [14]. It turned out that the matrix elements of the IR-field-induced radiative transitions are time-dependent. The used method was based on the Fourier series expansion of the time-dependent matrix elements. The first terms of these Fourier series, which are not time dependent, were only taken into account.

In the present paper we study the IR–field-induced single-photon generation by the current-carrying narrow thin-film ring, made of the isotropic *s*-pairing type-II superconductors such as *Nb₃Sn* and *V₃Ga*. For calculation of the matrix element for this problem we use a new method without Fourier series. The main result is that the boson-number distribution can be found by carrying out statistical measurements of the energies of photons emitted by the same ring.

## 2. Wave function in supercurrent state

We consider a flat thin-film ring made of the London superconductor with isotropic *s*-wave pairing. The film thickness, $d$, should be less than the IR field penetration depth for the superconductor. Then this field is homogeneous inside the film and has the same effect on all the Cooper pairs. To avoid absorption emitted photons in the ring material, the wall thickness of the ring is $b - a$ should be less than the photon absorption length. Here $a$ is the inner radius of the ring, $b$ is its outer radius which is assumed to be micron size.

In order to avoid excitations of fermionic quasiparticles and heating the film, we suppose that the ring temperature is much less than the temperature of superconducting transition, and the field frequency, $\omega_0$, is less than the doubled superconducting gap, $\hbar\omega_0 < 2\Delta$. Implying the *s*-wave superconductors such as *Nb₃Sn* with $\Delta \approx 5.3$ meV and *V₃Ga* with $\Delta \approx 4.9$ meV, the used radiation belongs to the far-infrared region. The magnetic field created by the supercurrent in the ring, is assumed to be small as compared with the first critical field for this type II superconductor. Then, the Abrikosov vortices are not formed that leads to a simpler form of the condensate wave function.

The wave function of a pair includes the wave function of the internal motion, wave function for the center of mass and the spin function of the pair. The first function has the characteristic length scale defined by the coherence length, which is $\approx 50$ angstroms for the abovementioned superconductors. For the supercurrent state in the ring the second function is represented by the phase function $\exp(i\phi_m)$.

The condensate wave function is anti-symmetric. For the singlet pairing it is achieved by its anti-symmetric spin part. Interaction of the condensate in the rings with electromagnetic fields does not affect the spin variables. Therefore, without any restriction, the spin wave function of the condensate can be omitted. Further, we assume that the wave functions describing the internal motion of the pairs do not change in the IR field. As a result, this phase function $\exp(i\phi_m)$ is only important for each boson, and for the ring considered, the condensate wave function (2) with $N$ bosons can be written as:

$$\Psi_{Nm}(\mathbf{\rho}) = \frac{1}{\sqrt{\Omega_r}} \exp(iN\phi_m), \qquad (3)$$

where $\Omega_r$ is the ring volume, $\phi_m = m\varphi$ is the phase, $m$ is the number of the magnetic induction flux quanta (fluxoids), trapped in the ring, $\varphi$ is the azimuthal angle in cylindrical coordinates tied to the ring ($a \leq \rho \leq b$ and $-d/2 \leq z \leq d/2$).

## 3. The phase in the electromagnetic field

$N$ and $\phi$ are canonically conjugate variables that implies $\Delta N \Delta \phi \geq 1/2$. In general, $N$ is large, and, hence, the phase may be treated quasiclassically [1]:

$$\frac{d}{dt}\hbar \nabla \psi = 2e\mathbf{F}_0 \cos(\omega_0 t), \qquad (4)$$

where $\psi$ is the field-induced change in the phase, and the right-hand side of (4) is the force on the Cooper pairs.

Let the wave vector of the coherent field irradiated the ring, is directed alone the $z$ – axis and the field amplitude, $\mathbf{F}_0$, is directed against the $y$ – axis. Then from Eq. (4) with the initial condition $\phi(t=0) = \phi_m$, we find the phase that describes the coherent oscillations of the condensate:

$$\phi_m(\mathbf{\rho},t) = m\varphi + \psi(\mathbf{\rho},t), \qquad (5)$$

with

$$\psi(\mathbf{\rho},t) = -\zeta_0 \rho \sin(\varphi) \sin(\omega_0 t), \qquad (6)$$

where $\mathbf{\rho}$ represents the condensate degrees of freedom, and $\zeta_0$ is the field parameter,

$$\zeta_0 = \frac{2eF_0}{\hbar \omega_0}. \qquad (7)$$

Using (4), we can introduce the potential energy of the condensate in the electromagnetic field, $U(\mathbf{\rho},t)$. Eq. (4) multiplied by $N$, gives us:

$$\nabla U(\boldsymbol{\rho},t) = -2e N \mathbf{F}_0 \cos(\omega_0 t) .$$

Then the condensate potential energy is

$$U(\boldsymbol{\rho},t) = -2e N \mathbf{F}_0 \boldsymbol{\rho} \cos(\omega_0 t) . \qquad (8)$$

Taking into account (5) and (6), in the external field the condensate wave function (3) is transformed as:

$$\Psi_{Nm}(\boldsymbol{\rho},t) = \frac{1}{\sqrt{\Omega_r}} \exp(iN(m\varphi + \psi(\boldsymbol{\rho},t))) . \qquad (9)$$

Now let us discuss restrictions on the parameter field (7). Eqs. (5) and (6) represent a collective mode of the coherent oscillations of condensate. This mode is induced by the alternating electric field of the incident IR wave. In order to the condensate oscillations would not lead to violation of the charge neutrality, the wave frequency must be high enough. From Eq. (6) we obtain the amplitude of these oscillations $\hbar \zeta_0 / m_C \omega_0$ ($m_C$ is the Cooper pair mass). The charge neutrality is conserved if this amplitude is very small compared with the coherence length. Then we have:

$$\zeta_0 \ll \frac{m_C \omega_0 \xi_0}{\hbar} .$$

As noted above, the field should not change the wave functions describing the internal motion of the pairs. Therefore the field energy of the pair $\mathbf{dF}_0$, where $\mathbf{d}$ is the pair dipole, should be small compared with the pair binding energy, $\Delta$. Using for estimation $d = e\xi_0$, where $\xi_0$ is the coherence length of the superconductor, we obtain:

$$\zeta_0 \ll \frac{\Delta}{\hbar \omega_0 \xi_0} .$$

Also, this field can change the ring energy in the initial states. To avoid this, we consider the field of low intensities, for which the field correction to the boson momentum $\hbar \nabla \phi_m$ is small. From (5) and (6) we get:

$$\zeta_0 \ll \frac{m}{b} . \qquad (10)$$

For the micron-sized ring made of the abovementioned superconductors and the IR field frequency $\omega_0 \propto 10^{12} s^{-1}$, the last restriction (9) on the field parameter is the most important one of the above findings.

It is important to find out how the energy ring is changed by the external field. In the London theory this energy is composed of the magnetic energy and the condensate kinetic energy,

$$E_m = \frac{\mu_0}{2} \int d\mathbf{r} \left( \mathbf{H}^2 + \lambda^2 \mathbf{j}^2 \right).$$

Here $\mathbf{j}$ is the current density and $\mathbf{H}$ is the magnetic field strength. In the state (9) the current is given by:

$$\mathbf{j} = \left( \Psi^*_{Nm}(\boldsymbol{\rho},t) \hat{\mathbf{j}} \Psi_{Nm}(\boldsymbol{\rho},t) \right) = \frac{2e\hbar n_C}{m_C} \left[ \frac{m}{\rho} \mathbf{i}_\varphi - \zeta_0 \sin(\omega_0 t) \mathbf{i}_y \right] - \frac{1}{\mu_0 \lambda^2} \mathbf{A}(\boldsymbol{\rho},t), \quad (11)$$

where $\mathbf{i}_\varphi$ the azimuthal unit vector, $\mathbf{A}$ is the vector potential, $\lambda$ is the London penetration depth. Considering conservation of the charge neutrality, the magnetic field strength is given by the equation: $rot\mathbf{H} = \mathbf{j} + \varepsilon_0 \partial \mathbf{F}/\partial t$. The displacement current can be neglected, since its ratio to the filed-induced oscillating current given by the second term in the square brackets in the right-hand side of (11), is equal to $(\lambda k)^2 <<< 1$, where $k$ is the IR field wave vector. Then, using $\mathbf{H} = \mu_0^{-1} rot\mathbf{A}$, the field correction to the ring energy is:

$$\delta E = N \frac{\hbar^2 \zeta_0^2}{2m_C} \sin^2(\omega_0 t) \left( 1 - \frac{1}{4\pi\lambda^2 \Omega_r} \iint \frac{d\mathbf{r} d\mathbf{r}_1}{|\mathbf{r}-\mathbf{r}_1|} \right).$$

Considering (10), we have $\delta E / E_m << m^{-2}$. Therefore, the field correction can be very small for the states with $m >> 1$. But more important is that this correction does not depend on the fluxoid number, and is the same for all current-carrying states of the ring. Hence, for the transition $m \to m_1$ the energy of the emitted photon does not include this filed correction, and is equal to $E_0(m^2 - m_1^2)$. Here $E_0 = \Phi_0^2/2L$ is the ring characteristic energy, $\Phi_0 = \pi\hbar/e$ is the fluxoid, $L$ is the ring self-inductance.

Thus, the potential energy (8) can be considered as perturbation that changes the phase (5) of the condensate wave function (9), whereas the field correction to the ring energies can be neglected.

## 4. Transition operator

The Hamiltonian of the system considered can be represented in the form:

$$H = H_0 + V + U. \quad (12)$$

Here, $H_0$ is the Hamiltonian for the initial supercurrent states with $N$ bosons and the free electromagnetic field. The first term of the interaction operator on the right-hand side of (12) is given by:

$$V = \int \mathbf{j}(N)\mathbf{A}_g d\mathbf{r}, \qquad (13)$$

where $\mathbf{j}$ is the operator of the superconducting current density in the ring, $\mathbf{A}_g$ is the operator of the vector potential of the electromagnetic field generated by the superconducting ring, and the integration is performed over the ring volume. The last term on the right hand side of (12) is the potential energy (8) of the condensate in the external field.

Because the ring energies remain unchanged in the field (8), the commutator of the potential energy of the Cooper pairs in the classical electromagnetic field $U$ and the Hamiltonian $H_0$ can be neglected. Then the evolution operator satisfies the equation:

$$i\hbar \frac{\partial}{\partial t} S(t,0) = \left[e^{iH_0 t} V e^{-iH_0 t} + U\right] S(t,0),$$

with $S(0,0) = 1$. Formal solution of this equation is:

$$S(t,0) = e^{iN\psi(t)} \left[1 - \frac{i}{\hbar} \int_0^t dt_1 e^{-iN\psi(t_1)} e^{iH_0 t_1} V e^{-iH_0 t_1} S(t_1,0)\right], \qquad (14)$$

where $\psi$ is given by (6).

Introducing the transition operator $T(t) = S(t,0) - e^{iN\psi(t)}$, from (14) we have for the single-photon generation:

$$T_1(t) = -\frac{i}{\hbar} e^{iN\psi(t)} \int_0^t dt_1 e^{-iN\psi(t_1)} e^{iH_0 t_1} V e^{-iH_0 t_1} e^{iN\psi(t_1)}. \qquad (15)$$

## 5. Single-photon generation

Using (15), the amplitude of the IR field-induced supercurrent transition $m \to m_1$ with the single-photon emission for the state with $N$ bosons can be presented in the form:

$$<\mathbf{k}, \Psi_{Nm_1} | T_1 | \Psi_{Nm}, 0> = -\frac{i}{\hbar} \int_0^t V_{m_1,m}^{\mathbf{k}}(t,t_1) \exp\left(-\frac{it_1}{\hbar}(E_m - E_{m_1} - \hbar\omega_k)\right) dt_1$$

with the matrix element of the interaction (12)

$$V_{m_1,m}^{\mathbf{k}} = \left(\frac{\hbar}{2\varepsilon_0 \omega_k}\right)^{1/2} \int \mathbf{l}_{\mathbf{k}\sigma} e^{-i\mathbf{k}\mathbf{r} + iN\psi(\boldsymbol{\rho},t)} \left(\tilde{\Psi}_{Nm_1}^*(\boldsymbol{\rho},t) \hat{\mathbf{j}} \tilde{\Psi}_{Nm}(\boldsymbol{\rho},t)\right) d\mathbf{r}, \qquad (16)$$

where the wave functions with the tilde sign are given by (9), $\mathbf{l}_{\mathbf{k}\sigma}$, $\mathbf{k}$ and $\hbar\omega_k$ are, respectively, the emitted photon polarization, wave vector and energy.

Using the well-known expression for the current density operator, the off-diagonal matrix element of the current density operator is reduced to:

$$\left(\Psi_{Nm_1}^* \hat{\mathbf{j}} \Psi_{Nm}\right) = \frac{e\hbar N}{m_C \Omega_r} \left\{\frac{m+m_1}{\rho} \mathbf{i}_\varphi - 2\zeta_0 \sin(\omega_0 t_1)\mathbf{i}_y\right\} e^{iN[(m-m_1)\varphi - \zeta_0 \rho \sin(\varphi)\sin(\omega_0 t_1)]}$$

$$-\frac{1}{4\pi\lambda^2}\int \frac{d\mathbf{r}_1}{|\mathbf{r}-\mathbf{r}_1|}\left(\Psi_{Nm_1}^*(\boldsymbol{\rho}_1, t_1)\hat{\mathbf{j}}\Psi_{Nm}(\boldsymbol{\rho}_1, t_1)\right). \tag{17}$$

The diamagnetic transition current represented by the second term on the right-hand side of (17), is suppressed since the integrand contains the very rapidly oscillating phase function. The second term in curly brackets of (17) can be omitted due to the restriction (10). Then the main contribution to the single-photon matrix element (17) is:

$$V_{m_1,m}^{\mathbf{k}} = \frac{2^{1/2}e}{m_C \Omega_r} * \left(\frac{\hbar^3}{\varepsilon_0 \omega_k}\right)^{1/2} * \frac{\sin\left(\frac{kd}{2}\cos\theta_k\right)}{k\cos\theta_k} I(t,N) \tag{18}$$

with

$$I(t,N) = N(m+m_1)\sin(\theta_l)\int_a^b d\rho \int_0^{2\pi} d\varphi \sin(\varphi_l - \varphi)$$

$$\exp\left(-ik_\rho \rho \cos(\varphi - \varphi_k) + iN[(m-m_1)\varphi - \zeta_0 \rho \sin(\varphi)\sin(\omega_0 t)]\right). \tag{19}$$

where $\theta_k$ and $\varphi_k$ are the polar and azimuthal angles of the photon wave vector, $k_\rho = k\sin\theta_k$, $\theta_l$ and $\varphi_l$ are the polar and azimuthal angles of the photon polarization $\mathbf{l}_{\mathbf{k}\sigma}$.

In contrast to the previous work [14], we use a new method for the analytical evaluation of the integrals (19). Note that the right-hand side of (19) is reduced to:

$$I = -N(m+m_1)e^{-iN(m-m_1)\varphi_g} \sin(\theta_l)\int_a^b d\rho \int_0^{2\pi} d\varphi \sin(\varphi - \varphi_l - \varphi_g)e^{iN(m-m_1)\varphi - iN\zeta_0 g(t)\rho\sin\varphi}, \tag{20}$$

where

$$g(t) = \left(\sin^2(\omega_0 t) + 2\frac{k_\rho \sin\varphi_k}{N\zeta_0}\sin(\omega_o t) + \left(\frac{k_\rho}{N\zeta_0}\right)^2\right)^{1/2}. \tag{21}$$

and

$$\sin\varphi_g = \frac{k_\rho \cos\varphi_k}{N\zeta_0 g(t)}. \tag{22}$$

Integrating over the azimuthal angle in (20), we obtain:

$$I(t,N) = i\pi N(m+m_1)e^{-iN(m-m_1)\varphi_g} \sin(\theta_l)\int_a^b d\rho$$

$$\left(e^{-i(\varphi_l+\varphi_g)}J_{N(m-m_1)+1}(N\zeta_0 g(t)\rho) - e^{i(\varphi_l+\varphi_g)}J_{N(m-m_1)-1}(N\zeta_0 g(t)\rho)\right), \tag{23}$$

where $J_{N(m-m_1)\pm 1}(N\zeta_0 g(t)\rho)$ is the Bessel function of the first kind with the very large order of $N(m-m_1)\pm 1$. This function does not vanish only when its argument is comparable with the order.

In (23) we pass to a new variable of integration $z$ defined by

$$z = 2^{1/3}(N(m-m_1)\pm 1)^{2/3}\left(\frac{\zeta_0 g(t)\rho}{m-m_1\pm N^{-1}}-1\right),$$

and use the well-known asymptotic expansion the Bessel functions with large orders in terms of the Airy function [15]:

$$J_{N(m-m_1)\pm 1}(N\zeta_0 g(t)\rho) = 2^{1/3}(N(m-m_1)\pm 1)^{-1/3} Ai(-z) + O\big((N(m-m_1))^{-1}\big).$$

Taking into account that the number of Cooper pairs $N \ggg 1$ in the superconductor, we obtain:

$$I(t) = 2\pi \frac{(m+m_1)}{\zeta_0 g(t)} e^{-iN(m-m_1)\varphi_g} \sin(\theta_l)\sin(\varphi_l + \varphi_g) \int_{A(t)}^{B(t)} Ai(-z)dz, \qquad (24)$$

where the time-dependent limits of integration ($A < B$), are

$$A(t) = 2^{1/3}(N(m-m_1))^{2/3}\left[\frac{\zeta_0 a g(t)}{m-m_1}-1\right] \qquad (25)$$

and

$$B(t) = 2^{1/3}(N(m-m_1))^{2/3}\left[\frac{\zeta_0 b g(t)}{m-m_1}-1\right]. \qquad (26)$$

Because the modulus of the matrix element (18) is of interest, the phase factor $e^{-iN(m-m_1)\varphi_g}$ in (24) can be omitted.

The limits of integration (25) and (26) can be large in the absolute value, and vary sharply with time. At non-positive values of $z$ the Airy function $Ai(-z)$ decreases exponentially with $|z|$. Here one third of the maximum value of the integral in right-hand side of (24) is accumulated in the region of $[-\eta_1, 0]$ with $\eta_1$ is of the order of several units. For the positive values $z$ the Airy function $Ai(-z)$ oscillates with $z$, and the oscillation amplitude decreases. Here two thirds of the maximum value of the integral in right-hand side of (24) is accumulated in the region of $[0, \eta_2]$ with $\eta_2$ is of the order of several tens. Outside the interval $[-\eta_1, \eta_2]$ the function $Ai(-z)$ is small. Therefore, if the integration region $[A(t), B(t)]$ does not include this interval $[-\eta_1, \eta_2]$, the integral on the right-hand side of (24) vanishes.

Because the number of the Cooper pairs in the ring $N \ggg 1$, the time-dependent limits of the integration (25) and (26) in the expression (24) intersect quickly the interval $[-\eta_1, \eta_2]$ with time variation. To estimate the intersection time, we consider the function (21) which should not be too small, otherwise both the lower and upper limits of integration are large and negative, and, hence, $I = 0$. Let the flat ring has the outer radius $b = 1\mu$, inner radius $a = b - \lambda/2$ and thickness $d = 0.1\lambda$, where the London penetration depth is taken $\lambda = 2*10^3$ angstroms as for $Nb_3Sn$ and $V_3Ga$. With the density of Cooper pairs $n_C = 10^{20} cm^{-3}$ in the superconductor, we obtain the number of Cooper pairs $\overline{N} = n_C \pi d(b^2 - a^2) \approx 1.2*10^6$.

According to (23) and (29), the upper limit of integration can be positive in certain time intervals if the field parameter $\zeta_0 > (m - m_1)/b$, where $m - m_1 \geq 1$ is the change of the fluxoid number upon the transition $m \to m_1$. Let $\zeta_0 = 2b^{-1}$. With the transition energy $E_m - E_{m_1} = 2eV$, we have $k_\rho \leq 10^5 cm^{-1}$ and the ratio:

$$\frac{k_\rho}{\overline{N}\zeta_0} \leq 4.2*10^{-6}.$$

Therefore, we can neglect the terms containing $k_\rho$ in the right-hand side of (21). This leads to:

$$g(t) = |\sin(\omega_0 t)|. \qquad (27)$$

Using (27), we can estimate the time at which the limits of integrations $A(t)$ and $B(t)$ pass through the interval $[-\eta_1, \eta_2]$:

$$\omega_0 \Delta t \propto \frac{\eta_1 + \eta_2}{\overline{N}^{2/3}} << 1.$$

Neglecting these transition regions with such small times and using (27), the expression (24) is reduced to the form:

$$I(t) = 2\pi \frac{m + m_1}{\zeta_0} \sin(\theta_l)\sin(\varphi_l)s(t), \qquad (28)$$

where

$$s(t) = |\sin \omega_0 t|^{-1} \theta(|\sin \omega_0 t| - \chi_b)\theta(\chi_a - |\sin \omega_0 t|). \qquad (29)$$

Here $\theta$ is the Heaviside step function, $\chi_b = \frac{m - m_1}{\zeta_0 b}$ and $\chi_a = \frac{m - m_1}{\zeta_0 a}$. Thus the photon generation stimulated by the far IR field, occurs only when

$$\frac{m - m_1}{\zeta_0 b} < g(t) < \frac{m - m_1}{\zeta_0 a}.$$

Taking into account this condition, from (22) we have

$$|\sin\varphi_g| < \frac{k_\rho b}{N(m-m_1)}.$$

For the ring considered, $|\sin\varphi_g| < 0.83*10^{-5}$, so that $\varphi_g = 0$ or $\pi$. Hence, in (24) we can replace $\sin(\varphi_l + \varphi_g) \to \sin(\varphi_l)$.

Note that the function $I(t)$ given by (28), does not depend on the number of Cooper pairs in the superconductor that consists with the result of [14]. Considering (19), we conclude that the perturbation of the superconducting condensate by the far IR electromagnetic field can effectively compensate the function $\exp(iN(m-m_1)\varphi)$ with very large phase. It leads to finite, really measurable lifetimes of the current states in superconducting thin-film rings of relatively small sizes, as shown below.

Using (1), the probability of the single-photon emission for the supercurrent decay channel $m \to m_1$ is written as:

$$w_{m_1 m}^s(t) = \frac{2\pi}{\hbar} \sum_{N,\mathbf{k}} c_N^2 \left\langle |V_{m_1 m}^{\mathbf{k}}|^2 \right\rangle_{pol} \delta(E_m - E_{m_1} - \hbar\omega_k), \qquad (30)$$

where $V_{m_1 m}^{\mathbf{k}}$ is the matrix element given by (18), $<...>_{pol} = \pi^{-1}\int_0^{2\pi} d\varphi_l ...$ means the average over the photon polarization.

From the equation $\mathbf{k}\mathbf{l}_{\mathbf{k}\sigma} = 0$ we have:

$$\sin(\theta_l) = \left(1 + tg^2(\theta_k)\cos^2(\varphi_l - \varphi_k)\right)^{-1/2} \qquad (31)$$

for the real polarization vectors. Hence, with account of (18) and (28) the average over the photon polarization in (30) is reduced to calculation of the function:

$$D(\theta_k, \varphi_k) = <\sin^2(\theta_l)\sin^2(\varphi_l)>_{pol} = \pi^{-1}\int_0^{2\pi} d\varphi_l \frac{\sin^2(\varphi_l)}{1 + tg^2(\theta_k)\cos^2(\varphi_l - \varphi_k)}.$$

Integrating over $\varphi_l$, we obtain:

$$D = (1 + \cos 2\varphi_k)|\cos\theta_k| - \frac{2\cos(2\varphi_k)}{tg^2\theta_k}(1 - |\cos\theta_k|). \qquad (32)$$

To sum over $N$ in (30), we note that the energy of the ring and, respectively, the characteristic energy $E_0$ are proportional to the Cooper pair number that is not fixed in the superconductor. Since $E_0 \propto R_L^{-1}$, where $R_L = L/\mu_0$ is the ring inductance in unit of $\mu_0$, we can introduce $R_L(N) = \bar{R}_L N/N$, where $\bar{R}_L$ corresponds to the average number of Cooper pairs. Also, the emitted photon energy depends on $N$ and, according to (30), is equal to $\hbar\omega_k(N) = \frac{N}{\overline{N}} E_0(\overline{N})(m^2 - m_1^2)$. Because the Cooper pair number is a fluctuating quantity, the emitted photon energy is fluctuating as well.

Using (32), the integration in (30) over the azimuthal angle $\varphi_k$ is easily carried out, and the probability of the single-photon emission is rewritten as:

$$w_{m_1m}^s(t) = \frac{2^5\alpha^2}{\pi}\left(\frac{2m_e}{mC}\right)^2 s^2(t) \frac{m+m_1}{m-m_1} \frac{\lambdabar_e^2 c \overline{R}_L}{d^2(b^2-a^2)^2 \zeta_0^2} F(\varepsilon_{mm_1}^d), \qquad (33)$$

where

$$F(\varepsilon_{mm_1}^d) = \sum_N \frac{\overline{N}}{N} c_N^2 \int_0^\pi \frac{\sin\theta_k d\theta_k}{|\cos\theta_k|} \sin^2\left(\frac{N}{\overline{N}}\varepsilon_{mm_1}^d \cos\theta_k\right). \qquad (34)$$

Here $\alpha$ is the fine structure constant, $m_e$ is the electron mass, $\lambdabar_e$ is the Compton wavelength of the electron, $c$ is the speed of light, and

$$\varepsilon_{mm_1}^d = \frac{dE_0(\overline{N})(m^2-m_1^2)}{2c\hbar}. \qquad (35)$$

It is easy to see from (34) that the energy distribution of photons emitted by the same ring for the same supercurrent decay channel $m \to m_1$, is a narrow peak determined mainly by the amplitudes $c_N$. This is due to the fluctuations of the number of Cooper pairs in the superconducting state. Passing from summation over $N$ to integration over the photon energy, the function (34) is reduced to $F = \int P(\hbar\omega_k) d\hbar\omega_k$, where $P(\hbar\omega_k)$ is defined by statistics of the emitted photon energy:

$$P(\hbar\omega_k) = \frac{\overline{N}}{\hbar\omega_k} c_{N=\frac{\overline{N}\hbar\omega_k}{E_0(\overline{N})(m^2-m_1^2)}}^2 \left[\gamma + \ln\left(\frac{d\hbar\omega_k}{c\hbar}\right) - Ci\left(\frac{d\hbar\omega_k}{c\hbar}\right)\right], \qquad (36)$$

where $\gamma$ is the Euler constant, $Ci$ is the integral cosine function, $c_N^2$ is the probability that the number of Cooper pairs is equal to $N$ in the superconducting ring.

As noted above, these amplitudes $c_N$ are not yet known. In [2,3] the wave packet (1) was formed with $c_N$'s given by the normal Gaussian distribution. With the standard deviation equal to $\sqrt{\overline{N}}$ the photon peak (36) is rewritten as:

$$P(\hbar\omega_k) = \sqrt{\frac{\overline{N}}{2\pi}} \frac{1}{\hbar\omega_k}\left[\gamma + \ln\left(\frac{d\hbar\omega_k}{c\hbar}\right) - Ci\left(\frac{d\hbar\omega_k}{c\hbar}\right)\right]\exp\left[-\frac{\overline{N}}{2}\left(\frac{\hbar\omega_k}{E_0(\overline{N})(m^2-m_1^2)}-1\right)^2\right]. \qquad (37)$$

According (37), the peak energy is $\hbar\omega_p = E_0(\overline{N})(m^2 - m_1^2)$ and FWHM of the peak is approximately equal to

$$\gamma = 2^{3/2} \frac{E_0(\overline{N})(m^2 - m_1^2)}{\sqrt{\overline{N}}}. \tag{38}$$

For the ring considered above, $\sqrt{\overline{N}} = 1.1*10^3$, and, as will be shown below, the photons with energies corresponded to the visible or near-UV light, can be emitted. We can therefore expect the peak widths of the order of a few meV that is favorable for the experimental observation.

The probability (33) is time dependent. Of course, the probability averaged over time, can be experimentally measured. Using the function (29), we obtain:

$$<w_{m_1 m}^s> = \frac{2^6 \alpha^2}{\pi^2} \left(\frac{2m_e}{m_C}\right)^2 \frac{\lambdabar_e^2 c R_L}{d^2 b^3 (1-(a/b)^2)^2} \frac{m+m_1}{(m-m_1)^2 \zeta_0} F(\varepsilon_{mm_1}^d) q_1(\chi_a, \chi_b), \tag{39}$$

where

$$q_1(\chi_a, \chi_b) = \begin{cases} (1-\chi_b^2)^{1/2}, & \chi_b < 1 \text{ and } \chi_a > 1 \\ (1-\chi_b^2)^{1/2} - \frac{a}{b}(1-\chi_a^2)^{1/2}, & \chi_a < 1 \end{cases}.$$

The time-averaged probability (39) defines the lifetime of the supercurrent for the decay channel $m \to m_1$ with the single-photon emission: $\tau_{m_1 m}^S = <w_{m_1 m}^S>^{-1}$.

## 6. Results and discussion

Upon receiving the results presented below, we take $B_{c1} = 20 \text{ mT}$ and $\lambda = 2*10^3 \, \text{Å}$ which are consistent with the data experimentally obtained for Nb$_3$Sn and V$_3$Ga [16]. The narrow thin-film ring has the outer radius $b = 1\mu$, inner radius $a = b - \lambda/2$ and thickness $d = 0.1\lambda$. The current density and magnetic induction distributions in the ring region were calculated by the use of method, presented in [17]. The magnetic field into the ring that is maximal at its inner edge should be less that the first critical field for the type II superconductor. Hence, the restriction on the fluxoid number is $m < m_c$, where $m_c$ is defined as: $B(\rho = a, m_c) = B_{c1}$. We obtained the following parameters of this ring: the characteristic energy $E_0 = 87$ meV, the inductance $L = 0.153$ nH and $m_c = 233$. Hereafter $m_C = 2m_e$ is used.

Fig. 1 shows the field parameter dependence of the time-averaged probability (39) of the single photon generation by the ring. Given the time-dependent functions $s(t)$ (29), the single-photon generation has a threshold dependence on the field parameter $\zeta_0$ (7). The generation threshold is defined by $\zeta_0^{th} = b^{-1}$. At $\zeta_0 < \zeta_0^{th}$ all the supercurrent transitions $m \to m_1$ are forbidden. It means that the supercurrent in the ring is persistent, and photon emissions do not occur.

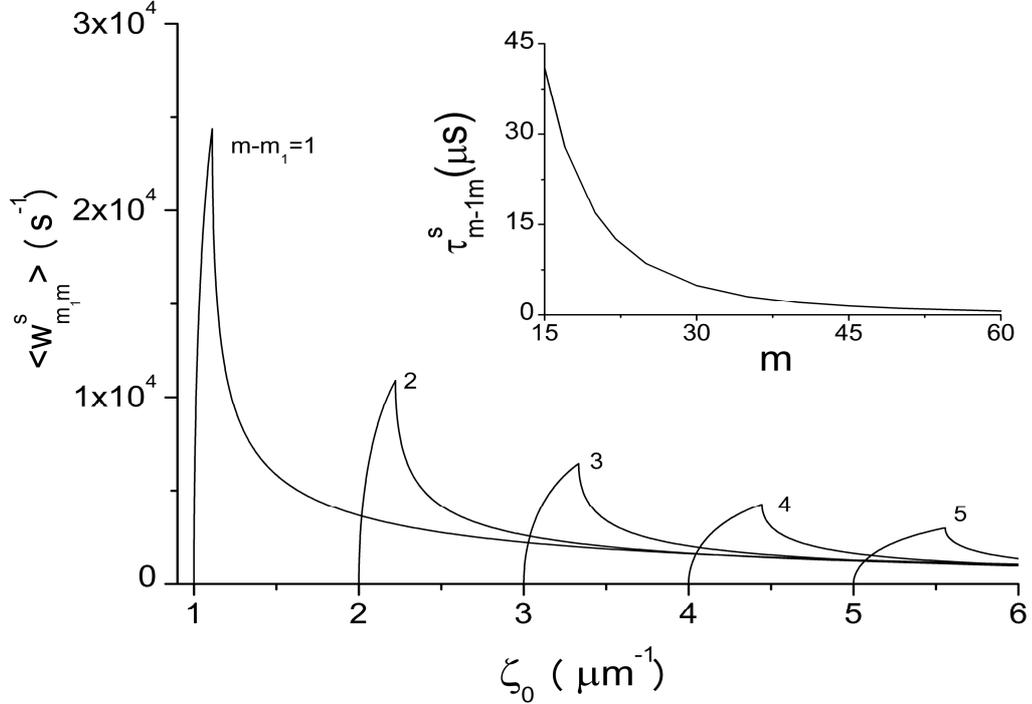

Fig. 1. The field parameter dependence of the time-averaged single-photon probability (39). The fluxoid number in the initial superconducting state is equal to $m = 15$. The peak 1 corresponds to destruction of one fluxoid in the final state ($m_1 = m - 1$); peak 2 - $m_1 = m - 2$; peak 3 - $m_1 = m - 3$; peak 4 - $m_1 = m - 4$ and peak 5 - $m_1 = m - 5$. Inset: the fluxoid number dependence of the minimum lifetime of the supercurrent for the decay channel $m \to m - 1$ with the single-photon emission.

The field parameter $\zeta_0$ (7) increases with increasing the IR field amplitude and decreasing its frequency. When $\zeta_0^{th} < \zeta_0 < 2\zeta_0^{th}$, the only supercurrent transition $m \to m - 1$ is allowed with destruction of one fluxoid in the final state of the superconducting ring. This transition is represented by the peak 1 in Fig. 1. For $m$ fluxoids in the initial state, the difference in energy between the initial and final states is equal to $E_0(N)(2m - 1)$. The shown peak corresponds to the photon energy $\hbar\omega_p = 2.52$ eV.

At $2\zeta_0^{th} < \zeta_0 < 3\zeta_0^{th}$ besides the already allowed transition $m \to m - 1$, the next supercurrent transition $m \to m - 2$, represented by the peak 2 in Fig. 1, becomes allowed. The photon energy for the second peak is $\hbar\omega_p = 4E_0(\overline{N})(m-1) = 4.87$ eV. When $n\zeta_0^{th} < \zeta_0 < (n+1)\zeta_0^{th}$, the number of the allowed supercurrent transitions is equal to $n$ with destruction of 1, 2, 3,... $n$ fluxoids in the final state (peaks 3,4 and 5 in Fig 1).

However, in our consideration the field parameter is restricted by (10), and, consequently, the number of transitions is limited well as. For the $m \to m-n$ transition the threshold intensity of the IR field, which increases quadratically with $n$, is equal to $n^2 I_{th}$, where

$$I_{th} = \frac{c\varepsilon_0 \hbar^2 \omega_0^2}{8e^2 b^2}.$$

Of course, the intensity must be kept low to avoid heating the superconducting film. Let us estimate this threshold intensity which corresponds to $\zeta_0^{th} = b^{-1}$. For the considered ring and the IR field frequency $\omega_0 = 10^{12} s^{-1}$ ($\hbar\omega_0 = 0.66$ meV $\ll \Delta$) we obtain $I_{th} = 14 \, mW/cm^2$ that is favorable for practical applications. The threshold intensity decreases with increasing the ring radius and decreasing the field frequency.

The maxima of the peaks presented in Fig. 1, correspond to the minimum lifetimes of the supercurrent transition $m \to m_1$ with the single photon emission. This lifetime decreases with increasing the number of fluxoids in the initial state (the inset to Fig. 1). At the same time the photon energies increase proportionally to $m$. Consequently, according to (38), the spectral width of the peak photon should increase well as. Therefore, in experiments it will be convenient to prepare the initial state of the rings with a large number $m$ which, however, is limited by $m_c$.

According (38), the ratio of the peak FWHM to the peak energy is

$$\frac{\gamma}{\hbar\omega_p} = \frac{2^{3/2}}{\sqrt{N}}. \qquad (40)$$

For experimental detection of the photon peak, its width $\gamma$ should be relatively large. This can be achieved in several ways. First, for a given initial state with the fluxoid number $m$, one can use the transition to a state with lower values of $m_1$. This is demonstrated in Fig. 2, in which the statistical distributions of the emitted photon energy by the same ring are presented for several transition channels $m \to m_1$. It is seen that the spectral widths increase with decreasing the fluxoid number in the final state of the superconducting ring. However, for excitation of the transition $m \to m-5$ the IR field intensity is approximately 25 times greater than the intensity for the transition $m \to m-1$. This way can be fraught with the problem of heating of the superconducting ring.

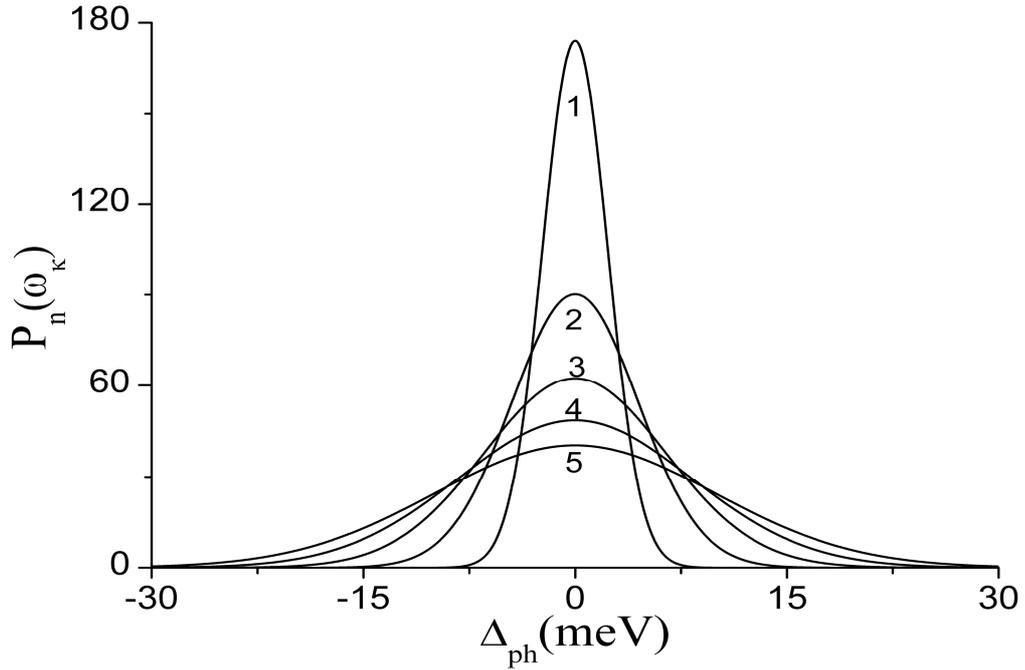

Fig. 2. The normalized photon spectrum (37), $P_n(\hbar\omega_k) = P(\hbar\omega_k)/\int P(\hbar\omega_k)d\hbar\omega_k$, as a function the photon energy relative to the peak energy, $\Delta_{ph} = \hbar\omega_k - \hbar\omega_p$. The parameter $m$ and label conventions are the same as for Fig. 1.

The second way of increasing the photon spectrum width is due to the preparation of the initial state of the ring with large fluxoid numbers and uses the IR field intensity at which the only transition $m \to m-1$ is allowed. In this case, the intensity is the lowest, $I_{th}$, and the width of the statistical distribution is $\gamma(meV) = 0.187*(2m-1)$, where $m < m_c = 233$.

Lastly, another possibility is associated with decreasing the average number $\overline{N}$ of Cooper pairs in the superconductor. Keeping unchanged the thickness and the ratio of the inner radius to the outer radius of the ring, it can be achieved using submicron-sized rings.

### 7. Conclusion

From a quantum field theoretical point of view, the spontaneous breaking of the gauge invariance means that the condensate state is in a coherent state which is the eigenfunction of the destruction field operator for the Cooper pairs, or a coherent superposition having definite phase and not definite number of the bosons.

Based on the presented method for the calculation of the time-dependent transition matrix elements without Fourier series, a theory of the single-photon emission from the thin-film superconducting rings irradiated by the far-IR electromagnetic fields was developed. The IR field causes the collective mode of the coherent oscillations of condensate that results to the radiative transition of all the Cooper pair involved to the supercurrent.

Uncertainty in the number of Cooper pairs in the superconducting states leads to the fact that statistical measurements of the energies of photons emitted by the same ring initially prepared in the same quantum state, give the photon peak. We have shown that experimental detection of this photon peak allows to find the boson-number distribution and to detail the coherent macroscopic quantum states of the superconducting condensate.

From a technological point of view, supercurrent-carrying thin-film rings irradiated by far IR fields, can be used as a new source of single photons in a wide range of their energies.

**Acknowledgments**
I am thankful to Prof. S.V. Sazonov for helpful discussions.